\documentstyle[12pt]{article}
\begin{document}

\begin{center}
{\Large \bf The Poincar\'{e} recurrence time for the de Sitter space with dynamical chaos}\\
\vskip .5cm
K. Ropotenko\\
\centerline{\it State Department of communications and
informatization} \centerline{\it Ministry of transport and
communications of Ukraine} \centerline{\it 22, Khreschatyk, 01001,
Kyiv, Ukraine}
\bigskip
\verb"ro@stc.gov.ua"

\end{center}
\bigskip\bigskip

\begin{abstract}
For an ordinary thermodynamical system the Poincar\'{e} recurrence
time is exponentially large in the Boltzmann entropy of the system.
It turns out, that for a system with dynamical chaos it is
determined by the Kolmogorov-Sinai entropy and can be considerably
shorter. It is shown in this note that for the de Sitter space with
strong mixing properties the mean recurrence time is equal to the
inverse Hubble constant. This means that our universe can have a
finite lifetime bounded by the current age of the universe. After
this time, the universe should recycle itself and this process has
neither a beginning nor an end.
\end{abstract}
\bigskip\bigskip

Recent observations suggest that our universe most likely tends to
the de Sitter state \cite{Ast}. In accordance with the inflationary
cosmology, a similar state existed also at the beginning
approximately $13.7$ billions years ago ($\sim 10^{17}\rm s$). In
this connection, the important questions arise: Is there a relation
between these states, and Is the number $10^{17}\rm s$ arbitrary, or
not? A possible answer proposed by Dyson, Kleban and Susskind (DKS)
in \cite{DKS} is to regard the early universe as a rare
thermodynamical fluctuation of the present de Sitter state taken as
equilibrium, due to the Poincar\'{e} recurrence. According to the
Poincar\'{e} recurrence theorem \cite{DKS}, any state of an isolated
finite system continuously returns arbitrarily close to its initial
value in a finite amount of time (the Poincar\'{e} recurrence time,
$t_{r}$). For an ordinary thermodynamical system this time is
exponentially large in the number of elements of the system, or in
the thermodynamical (Boltzmann) entropy of the system, $S$:
\begin{equation}
\label{time}t_{r}\sim e^{S}.
\end{equation}

As is well known, de Sitter space is a thermodynamical system with
the Gibbons-Hawking (GH) entropy of just the same kind given by
\begin{equation}
\label{GH}S_{GH}=\frac{A}{4 l_{P}^{2}}=\frac{\pi}{H^{2} l_{P}^{2}
}\,,
\end{equation}
where $H$ is the Hubble constant, and the area of the event horizon
$A$ is related with the radius of de Sitter space $R_{dS}$,
$R_{dS}=H^{-1}$, in the usual way $A=4\pi R_{dS}^{2}$.

Since the GH entropy of the present de Sitter state is of order
$10^{120}$, it follows that the Poincar\'{e} recurrence time for the
fluctuation to occur is
\begin{equation}
\label{tim}t_{r}\sim \exp (10^{120})
\end{equation}
that is considerably larger than $10^{17}\rm s$. So the current age
of the universe cannot be regarded as the Poincar\'{e} recurrence
time but rather as a relaxation time. Usually the relaxation time is
a very complicated phenomenological quantity depending on many
factors. But the age of the universe $t_{0}\sim 10^{17}\rm s $ is
related with such a fundamental quantity as the radius of de Sitter
space, $t_{0}\sim H_{0}^{-1}$, and thus it cannot be a
phenomenological quantity. Can the Poincar\'{e} recurrence time be
much shorter than (\ref{tim})? Moreover, can the Poincar\'{e}
recurrence time be equal to $H_{0}^{-1}$?

In this note, I want to point out to a possible relation between the
Poincar\'{e} recurrence time and the radius of the de Sitter space
with dynamical chaos.

We begin with definitions. One of the most important quantities
characterizing the chaotic behavior of a dynamical system is the
Kolmogorov-Sinai (KS) entropy, which describes the rate of change of
information about the phase space trajectories as a system evolves.
Suppose that phase space of a dynamical system is finite, then the
KS entropy $h_{KS}$ is the sum of all the positive Lyapunov
exponents of the system, where the Lyapunov exponents $h$
characterize the rate of exponential separation of the nearby
system's trajectories in phase space due to a local instability
\cite{zas}

\begin{equation}
\label{tr} d(t)=d(0)\:e^{ht}.
\end{equation}
This leads to an increase in the phase space volume occupied by the
system with time
\begin{equation}
\label{ph} \Delta{\Gamma(t)}=\Delta{\Gamma(0)}\:e^{h_{KS}t}.
\end{equation}
As a result the thermodynamical entropy increases
\begin{equation}
\label{B} S(t)=h_{KS}t+\ln(\Delta{\Gamma(0)}).
\end{equation}
As is easily seen, the KS entropy $h_{KS}$ is not really an entropy
but an entropy per unit time, or entropy rate, $dS/dt$.

It is well known that general relativistic systems described by the
Einstein equations can exhibit chaotic behavior \cite{hob}. Susskind
has shown \cite{s} that stringy matter near the event horizon of a
black hole with the gravitational radius $R_{g}$ exhibits
instability (the spreading effect): a string approaching the event
horizon spreads in the transfers directions over the horizon in the
reference frame of an external observer like $\label{rad}\langle
R_{s}\rangle ^{2}\sim e^{t/R_{g}}$. But this means nothing but that
close trajectories of bits of the string diverge exponentially as in
(\ref{tr}). So, taking into account (\ref{ph}) and (\ref{B}), we can
attribute to the black hole the KS entropy \cite{Rop},
\begin{equation}
\label{KS} h_{KS} = \frac{1}{R_{g}}.
\end{equation}

Obviously, the same phenomenon will be also observed in the de
Sitter case: string matter approaching the event horizon of de
Sitter space spreads over the horizon. Thus, repeating our
arguments, we can obtain the KS entropy for the de Sitter space
\cite{Rop}
\begin{equation}
\label{KSG} h_{KS}= H.
\end{equation}

The KS entropy measures the rate at which information about the
string state is lost as the string spreads over the horizon. Since
the entire accessible phase space of the string is bounded by the
horizon area, the trajectories of the string mix together and the
mixing time will be finite, $\sim H^{-1}$.

We have obtained the KS entropy for the de Sitter space by means of
a string spreading over the event horizon. It is widely believed,
however, that the spreading effect is not a peculiar feature of a
special (still hypothetical) kind of matter. It turns out that in
the framework of the so-called infrared/ultraviolet connection
\cite{s} it is a general property of all matter at energies above
the Planck scale.

Finally, let us turn to the Poincar\'{e} recurrence time. It should
be emphasized that the Poincar\'{e} recurrence theorem has nothing
to do with appearance of statistical properties in a system.
Recurrences exist in both quasiperiodical and stochastic motions.
Moreover the Poincar\'{e} recurrence theorem does not say anything
about the mean recurrence time, $\langle t_{r} \rangle$. Chaotic
dynamics provides a new approach to the determination of the mean
recurrence time by means of the distribution of recurrence times,
$f(t_{r})$. The properties of $f(t_{r})$ has been studied for a
number of models of dynamical chaos and it was found that (see
\cite{zasl} and references therein)
\begin{equation}
\label{timee}f(t_{r})=(1/\langle t_{r} \rangle ) \exp (-t/\langle
t_{r}\rangle).
\end{equation}
It turns out that the stronger the local instability the shorter
$\langle t_{r}\rangle$; in systems with strong mixing properties
(with the exponential divergence (\ref{tr}) and a finite mixing
time) the mean recurrence time is
\begin{equation}
\label{timee}\langle t_{r}\rangle =\frac{1}{h_{KS}}.
\end{equation}
Hence, taking into account (\ref{KSG}), we obtain
\begin{equation}
\label{timee}\langle t_{r}\rangle = H^{-1},
\end{equation}
as required. In other words, the Poincar\'{e} recurrence time
coincides with the age of the universe for the de Sitter space with
dynamical chaos.

From this an important conclusion follows: if our universe is in a
de Sitter state with strong mixing properties, its lifetime will be
finite and bounded by the mean recurrence time, $\langle
t_{r}\rangle \sim 10^{17}\rm s$. After this time, the universe
should recycle itself. This process has neither a beginning nor an
end.

\end{document}